          [2001/04/25 1.1 (PWD)]
\documentclass[a4paper,twocolumn]{esapub2005} % European paper
\usepackage{times}
\usepackage{natbib}
\usepackage{graphicx}

\newcommand{\inte}{\textsl{INTEGRAL }}
\newcommand{\xte}{\textsl{RXTE }}
\newcommand{\asm}{\textsl{ASM }}
\newcommand{\ibis}{\textsl{IBIS (ISGRI) }}
\newcommand{\jemx}{\textsl{JEM-X }}
\newcommand{\spi} {\textsl{SPI}}
\newcommand{\xspec}{\textsl{XSPEC}}
\newcommand{\ttm}{\textsl{TTM $\&$ HEXE}}
\newcommand{\osse}{\textsl{OSSE}}
\newcommand{\cgro}{\textsl{CGRO}}

\title{Pulse Period History and Phase Resolved Spectra of 1A 0535+262}
\author[1]{I.Caballero}
\author[2]{ P. Kretschmar}
\author[1]{A.Santangelo}
\author[1,3]{A.Segreto}
\author[3]{C.Ferrigno}
\author[1]{R.Staubert}
\affil[1]{Institut f\"ur Astronomie und Astrophysik, Universit\"at T\"ubingen, Germany}
\affil[2]{European Space Astronomy Center, ESA, Madrid, Spain}
\affil[3]{Instituto di Astrofisica Spaziale (IASF-INAF), Palermo, Italy}
\begin{document}


\maketitle

%%------------------------------------------------------------------------------------
%%					ABSTRACT			
%%------------------------------------------------------------------------------------
\begin{abstract}
The Be/X-ray binary 1A 0535+262 was discovered in 1975 during a
giant outburst. Afterwards it has shown periods of quiescence (flux below 10
mCrab), normal outbursts (10 mCrab-1Crab) and occasionally giant outbursts
(several Crab). Ending 11 years of quiescence, the
last giant outburst took place in May/June 2005, but the source was
too close to the Sun to be observed by most satellites. A subsequent
normal outburst took place in August 2005, which was observed by \inte\ and
\xte\ TOO observations. Based on \inte\ data, we present results on the 
long term pulse 
period history of the source, on their energy dependent pulse profiles and on phase 
resolved spectra.
\end{abstract}
%%------------------------------------------------------------------------------------
%%------------------------------------------------------------------------------------

%%------------------------------------------------------------------------------------
%%					INTRODUCTION			
%%------------------------------------------------------------------------------------
\section{Introduction}
\begin{figure}
\centering
\includegraphics[width=0.9\linewidth]{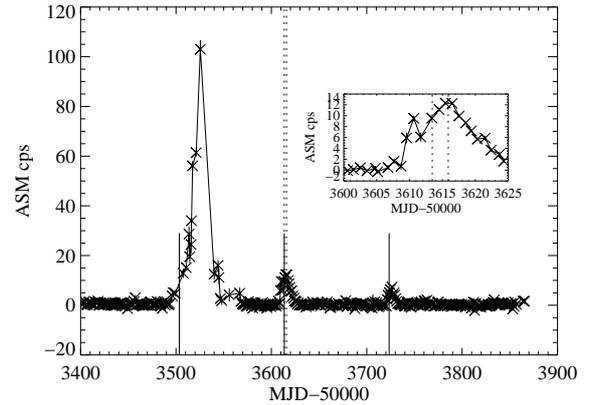}
\caption{\xte \asm long term light curve of the last three outbursts. 
The dotted lines indicate the time of our \inte observation, 
the inset gives a blow-up. The vertical lines indicate 
the times of periastron passage.\label{fig:asm_lc}}
\end{figure}

Since its discovery in 1975 \cite{rosenberg75}, the Be/X-ray binary 1A
0535+262 has been intensely studied. Further details may be found in a review
by \cite{giovannelli92}. The source has shown giant outbursts in April/May
1975 \cite{rosenberg75}, October 1980 \cite{nagase82}, March/April 1989
\cite{sunyaev82}, February 1994 \cite{finger94}, and in May/June
2005 \cite{tueller05}. Then the source showed a normal outburst 
in August/September 2005,
observed by \inte and \xte. During this outbusrt, the averge flux was 300
mCrab in the energy range 5-100 keV \cite{kretsch05}. The last normal 
outburst took place in December 2005 \cite{finger05}. 
Fig.~\ref{fig:asm_lc} shows the \xte \asm long term light curve 
of the source during the last three outbursts. 
In this paper, we present preliminary results on the timing and 
spectral behaviour of the source based on \inte \ibis data from the 
August/September 2005 outburst. All data were reduced 
and analysed using \inte OSA v5.1. Pulse phase resolved spectroscopy 
was performed using additionally the software provided by IASF/INAF Palermo.
%%------------------------------
%%------------------------------
%%------------------------------------------------------------------------------------
%%------------------------------------------------------------------------------------

%%------------------------------------------------------------------------------------
%%			PULSE PERIOD HISTORY AND PULSE PROFILES			
%%------------------------------------------------------------------------------------
\section{Pulse period history and pulse profiles}
\begin{figure}
\centering
\includegraphics[angle=90,width=0.90\linewidth]{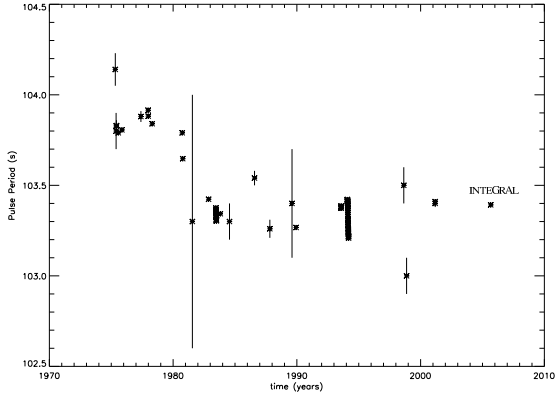}
\caption{Pulse period history of 1A 0535+262.\label{fig:period_history}}
\end{figure}

\begin{figure}[h]
\centering
\includegraphics[width=1.3\linewidth]{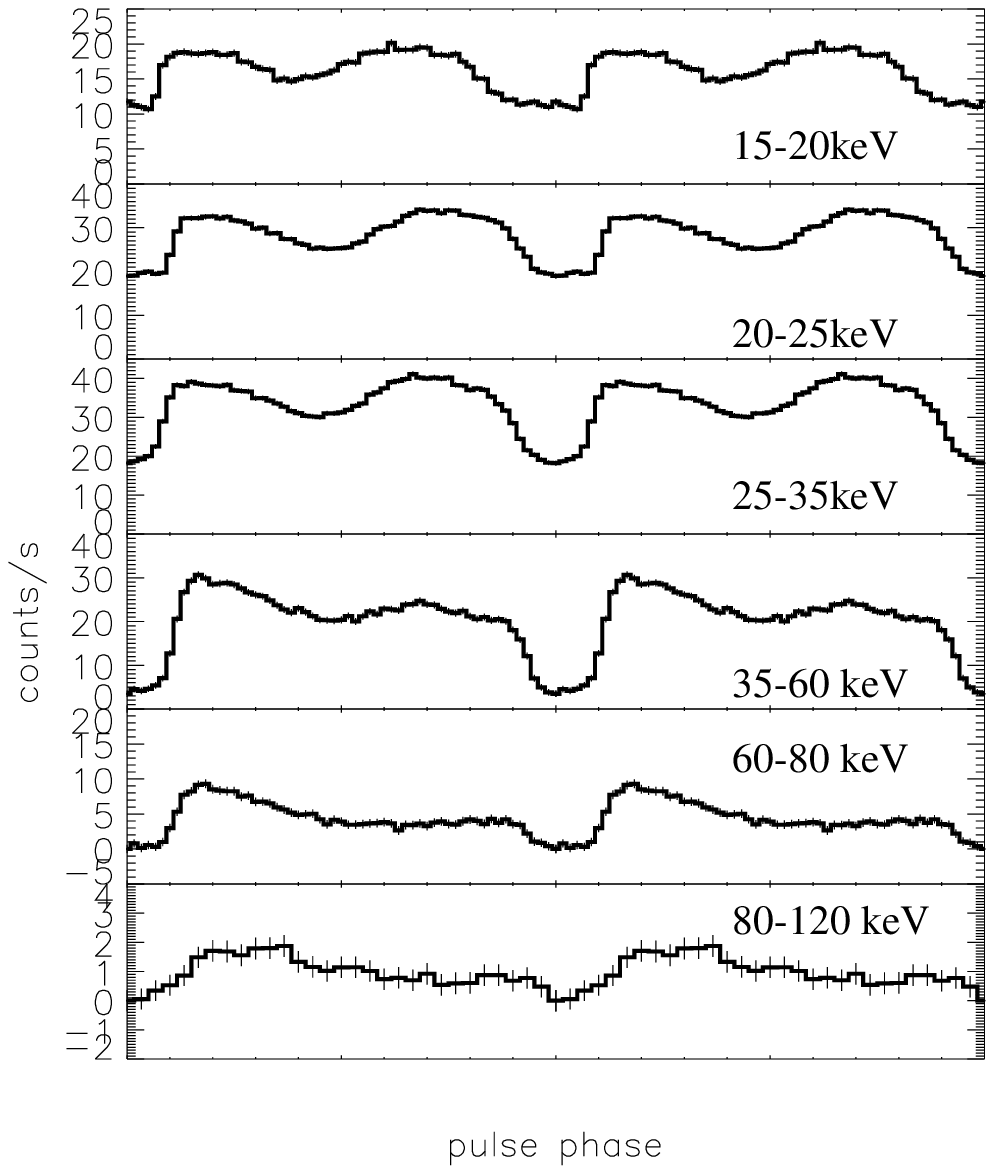}
\caption{1A 0535+262 \ibis pulse profiles.\label{fig:ibis_pp}}
\end{figure}

Using epoch folding techniques, we calculated the pulse period of the 
source, after applying barycentric correction and a correction for 
the binary orbit. The orbital parameters were taken from \cite{finger96}
and \cite{coe06}. In order to find a $\dot{P}$ , we divided the observation into 
27 intervals of $\sim$ 6ks each. We folded the light curve of each of 
those intervals over the period obtained from epoch folding, and 
checked that the pulse profiles obtained are shifted by an integer number 
of periods. We find a constant period of P=103.3920 $\pm$ 0.0004s for MJD 53613.460475.
For comparison, in Fig.~\ref{fig:period_history} we plot the pulse 
period history of the source since its first determination in 1975.
Using the above pulse period obtained, we folded the light curves 
for \ibis data in different energy ranges. The resulting  pulse 
profiles are plotted in Fig.~\ref{fig:ibis_pp}. Two pulse phases are shown for clarity. 
A double peak pattern is seen up to at least 60\,keV, while at 
higher energies one of the peaks appears to be strongly reduced. 
The source is observed to pulsate up to $\sim120$~keV. It is
evident form Fig.~\ref{fig:ibis_pp} the strong variation of the pulsed
fraction with the energy. 
%, while above 120 keV 
%no significant flux is detected, but apparently no modulation is detected.

%%------------------------------
%%------------------------------
%%------------------------------
%%------------------------------------------------------------------------------------
%%------------------------------------------------------------------------------------

%%------------------------------------------------------------------------------------
%%				PHASE AVERAGE SPECTRA			
%%------------------------------------------------------------------------------------
\section{Phase average spectra}
\begin{figure}[h]
\centering
\includegraphics[angle=90,width=1.0\linewidth]{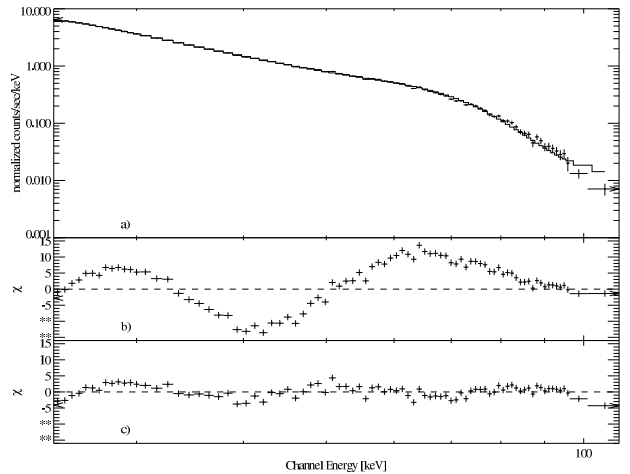}
\caption{(a) \ibis phase average spectra and model (powerlaw with 
a high energy cutoff and two cyclotron lines at $\sim$ 45 keV and 
$\sim$ 100 keV), (b) residuals for a fit without cyclotron lines 
($\chi^2_{\mathrm{red}}$=32.138/132dof) (c) residuals for a fit 
including two cyclotron lines ($\chi^{2}_{\mathrm{red}}$=1.7/126dof).
\label{fig:phase_average}}
\end{figure}

We extracted the phase average spectra for \ibis. To model the continuum we use a 
powerlaw with a high energy cutoff (\xspec\ HighECut), freezing the photon index 
to $\alpha \sim 1$ and the cutoff energy to $E_c\sim 20$~keV (theses values, 
typical for accreting pulsars, were taken from the literature, 
see \cite{kend94}). When fitting this continuum we detected two absorption 
like features in the residuals, that we interpret as cyclotron resonance 
scattering features. We modeled those features using Gaussian lines in absorption
\cite{coburn02}. The $\chi^2_{\mathrm{red}}$ obtained for a fit without 
absoprtion lines is 32.14 for 132 d.o.f. Including one Gaussian 
absorption line at $\sim45$keV the $\chi^2_{\mathrm{red}}$ improves to 
3.315 for 129 d.o.f. By including two Gaussian absoprtion lines at 
$\sim45$keV and $\sim100$keV, the $\chi^2_{\mathrm{red}}$ further 
reduces to 1.7 for 126 d.o.f. We therefore included in our model two 
absorption lines (see Fig.~\ref{fig:phase_average}). The best fit 
parameters for the lines are listed in Table~\ref{tab:table1}. 
%%------------------------------
\begin{table}[h!]
  \begin{center}
    \caption{Parameters of the cyclotron lines obtained from the phase average spectra.
		 Uncertainties are 90 \%  confidence for one parameter of interest
                  ($\chi^{2}_{min}+2.706$).}\vspace{1em}
    \renewcommand{\arraystretch}{1.0}
    \begin{tabular}[h]{lc}\hline
	     $ E_{1} (keV)      $  &  $   46.1 {+0.5 \atop -0.5}      $      \\\hline
	     $ \sigma_{1}(keV)  $  &  $    8   {+1 \atop - 3}         $      \\\hline
	     $ \tau_{1}         $  &  $    0.35{+0.03 \atop -0.01}    $      \\\hline
	     $ E_{2}(keV)       $  &  $   106   {+7 \atop -4}          $      \\\hline
	     $ \sigma_{2}(keV)  $  &  $    12   {+3 \atop -2}          $      \\\hline
	     $ \tau_{2}         $  &  $    1.1 {+0.3 \atop -0.2}      $      \\\hline
	     $ \chi^2_{\mathrm{red}}$/dof  &  $    1.7/126 			      $      \\\hline    	
    \end{tabular}
    \label{tab:table1}
  \end{center}
\end{table}
%%------------------------------
%%------------------------------
%%------------------------------
%%------------------------------------------------------------------------------------
%%------------------------------------------------------------------------------------

%%------------------------------------------------------------------------------------
%%				PHASE RESOLVED SPECTRA			
%%------------------------------------------------------------------------------------
\section{Phase resolved spectra}
\begin{figure}[h]
\centering
\includegraphics[angle=90,width=0.90\linewidth]{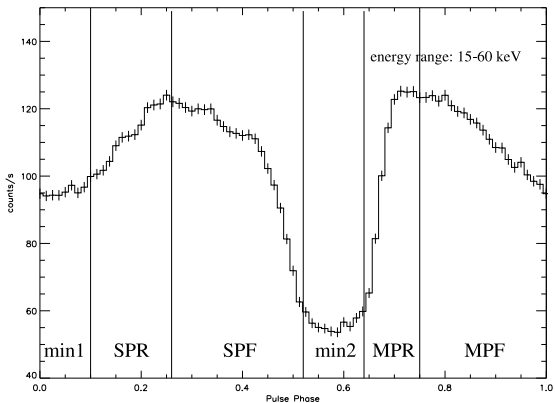}
\caption{phase selection for phase resolved spectroscopy.\label{fig:phase_selection}}
\end{figure}

\begin{figure}[h]
\centering
\includegraphics[bb=77 38 573 711,clip=,angle=270,width=1.0\linewidth]{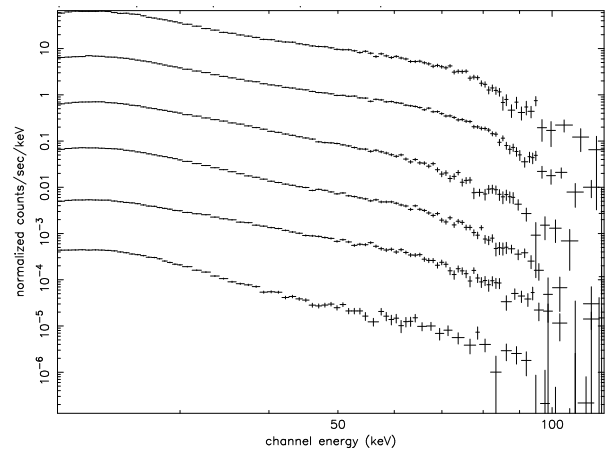}
\caption{Phase resolved spectra for all phases (from top spectrum to 
bottom, corresponds to 'MPR', 'MPF', 'SPR', 'SPF', 'min1', `min2', with 
normalisation factors for plotting 10, 1, $10^{-1}$, $10^{-2}$, $10^{-3}$, 
$10^{-4}$ respectively). 
A clear variation with phase is seen.\label{fig:phase_resolved}}
\end{figure}

To perform phase resolved spectroscopy we divided
the pulse profile into six phase intervals: main peak rise (MPR), main peak fall
(MPF), secondary peak rise (SPR), secondary peak fall (SPF), and the two
minima (min1, min2) (Fig.~\ref{fig:phase_selection}). \ibis spectra were 
accumulated for each phase interval (Fig.~\ref{fig:phase_resolved}).
To model the continuum, we again used a powerlaw with a high energy 
cutoff ( \xspec\ HighECut) in the same way as we did for phase average 
spectroscopy. We clearly detect one cyclotron resonance scattering feature at 
$\sim$ 45 keV in the residuals of all phase intervals. Furthermore, we find 
in most phases ('SPR', 'SPF', 'MPR', 'MPF') a second cyclotron line at $\sim$ 100 keV. 
%%------------------------------
%%------------------------------
%%------------------------------
%%------------------------------
For the phase intervals 'SPR', 'SPF', 'MPR'  we freeze the energy of the 
second line to 100 keV for the fitting. For the phase interval 'MPF' we can 
fit the two lines well. The best fit parameters are listed in Table~\ref{tab:table2}.
%%------------------------------
\begin{table*}
  \begin{center}
    \caption{Best fit parameters for the cyclotron lines. From these 
	     preliminary results, the centroid
	     of the fundamental cyclotron line seems to be rather 
	     constant within the phase.  However the width and depth of the 
	     line changes with phase.}\vspace{1em}
    \renewcommand{\arraystretch}{1.0}
    \begin{tabular}[h]{lcccccc}\hline
	& \textbf{min1}& \textbf{SPR} & \textbf{SPF} & \textbf{min2}& \textbf{MPR} & \textbf{MPF} \\[1ex] \hline\hline\\

	$E_{1} (keV)      $ &   $47.2{+0.9 \atop -0.9}   $  & $47.5 {+0.9 \atop -0.8}  $  & $45.9 {+0.6 \atop -0.}  $     &
	                        $45.7 {+1.3 \atop -1.2}   $  & $44.3 {+0.9 \atop -0.7}  $  & $45.8 {+0.6 \atop -0.5}  $     \\\hline
	$\sigma_{1} (keV) $ &   $ 5.0 {+0.9 \atop -0.8}   $  & $ 7.1 {+1.0 \atop -0.9}  $  & $ 8.8 {+0.7 \atop -0.7}  $     &
			        $ 7.8 {+0.9 \atop -0.9}   $  & $ 9.5 {+0.9 \atop -0.8}  $  & $ 9.5 {+0.7 \atop -0.6}  $     \\\hline
	$\tau_{1}         $ &   $0.22 {+0.03 \atop -0.03} $  & $ 0.34{+0.05 \atop -0.04}$  & $ 0.58{+0.07 \atop -0.04}$     &
				$0.70 {+0.09 \atop -0.09} $  & $ 0.36{+0.03 \atop -0.03}$  & $0.36 {+0.03 \atop -0.02}$     \\\hline
	$E_{2} (keV)      $ &   $            -            $  & $           100          $  & $           100          $     &
			        $            -            $  & $      100        $  & $102  {+4 \atop -3}      $     \\\hline
	$\sigma_{2} (keV) $ &   $            -            $  & $            24  {+4 \atop -3}         $  & $ 27  {+3 \atop -2}                  $     &
                                $            -            $  & $10 {+2 \atop -2}   $  & $10 {+3 \atop -2}  $     \\\hline
	$ \tau_{2}        $ &   $            -            $  & $1.1 {+0.3 \atop -0.2}   $  & $ 1.6 {+0.3 \atop -0.2}  $     &
			        $            -            $  & $1.3 {+0.6 \atop -0.3}   $  & $ 1.4 {+0.4 \atop -0.3}  $     \\\hline
	$ \chi_{red}^{2}/$dof      &   $          1.09/129       $  & $        1.02/127        $  & $      1.4/127          $     &
				$          1.4/129        $  & $        1.40/127         $  & $      1.09/126          $     \\\hline

     \end{tabular}
     \label{tab:table2}
  \end{center}
\end{table*}
%%------------------------------
%%------------------------------
\begin{figure}
\centering
\includegraphics[angle=90,width=1.0\linewidth]{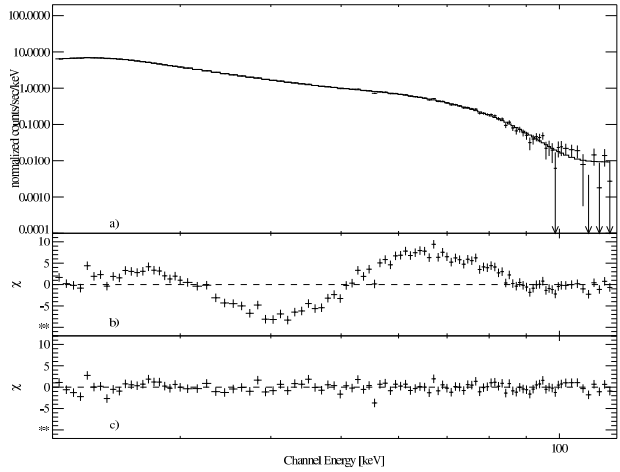}
\caption{(a) \ibis spectrum and model and for the main peak fall 'MPF' 
(power law with a high energy cutoff and two cyclotron lines at  
$\sim45$~keV and $\sim100$~keV). (b) residuals for a fit without 
the cyclotron lines. (c) residuals for a fit including two cyclotron 
lines ($\chi^{2}=1.16$).\label{fig:MPF_new_plot}}
\end{figure}
%%------------------------------
Fig.~\ref{fig:MPF_new_plot}(a) shows the spectrum and model for the 
main peak fall 'MPF' (powerlaw with a high energy cutoff and two 
cyclotron lines at $\sim45$~keV and $\sim100$~keV). 
Fig.~\ref{fig:MPF_new_plot}(b) shows the residuals for a fit without 
the cyclotron lines.  Fig.~\ref{fig:MPF_new_plot}(c) shows the 
residuals for a fit including two cyclotron lines ($\chi^{2}=1.16$). 
%%------------------------------------------------------------------------------------
%%------------------------------------------------------------------------------------

%%------------------------------------------------------------------------------------
%%					SUMMARY			
%%------------------------------------------------------------------------------------
\section{Summary}
In this poster we show some first results of our analysis of the \inte 
1A 0535+262 data from the August/September 2005 outburst. The source 
is found to pulsate up to 120 keV and energy dependent pulse profiles 
are observed. We also detect the presence of two phase dependent 
cyclotron lines at $\sim$~45~keV and $\sim$~100~keV in the hard X-ray spectrum 
of the source. This therefore confirms previous results by 
Kendziorra et al. \cite{kend94} based on \ttm\ data taken during the 
March/April 1989 giant outburst. We wish to outline however that 
observation from  \osse\  on \cgro\ during the February 1994 giant outburst 
clearly detected only the 110 keV line. The presence of the fundamental line 
was not clear, but it was concluded that if the line was present, 
its optical depth should be significantly smaller than that of the 
110 keV line \cite{grove95}. Based on our results we can estimate from the 
fundamental line at $\sim45$~keV the magnetic field of the source to 
be $\sim4$~x$10^{12}$~G. Further detailed analysis of this outburst is ongoing. 
The next step will be to try different models for fitting the broad 
band continuum making use of observational data from \jemx\ and \spi. 
%%------------------------------------------------------------------------------------
%%------------------------------------------------------------------------------------

%%------------------------------------------------------------------------------------
%%					BIBLIOGRAPHY		
%%------------------------------------------------------------------------------------

%%------------------------------------------------------------------------------------
%%------------------------------------------------------------------------------------
\end{document}